\documentclass[fleqn,usenatbib]{mnras}
\usepackage{amsmath}
\usepackage{graphicx}
\usepackage{amssymb}
\usepackage{color}
\usepackage[caption=false]{subfig}
\usepackage{natbib} 
\usepackage{url} 
	
\usepackage[capitalise]{cleveref}
\newcommand{\blind}{0}

\begin{document}

 \title{How to estimate Fisher information matrices from simulations}
 \author[Coulton and Wandelt.]{
William R. Coulton$^{1}$\thanks{E-mail:william.coulton@gmail.com)}
and Benjamin D.~Wandelt$^{2,1}$
\\
$^{1}$Center for Computational Astrophysics, Flatiron Institute, 162 5th Avenue, New York, NY 10010, USA\\
$^{2}$Sorbonne Universit\'{e}, CNRS, UMR 7095, Institut d'Astrophysique de Paris, 98 bis bd Arago, 75014 Paris, France\\
}


\if0\blind
{
 \maketitle
} \fi

\if1\blind
{
 \bigskip
 \bigskip
 \bigskip
 \begin{center}
 {\LARGE\bf How to estimate Fisher information matrices from simulations}
\end{center}
 \medskip
} \fi

\bigskip
\begin{abstract}
The Fisher information matrix is a quantity of fundamental importance for information geometry and asymptotic statistics. In practice, it is widely used to quickly estimate the expected information available in a data set and guide experimental design choices. In many modern applications, it is intractable to analytically compute the Fisher information and  Monte Carlo methods are used instead. The standard Monte Carlo method produces estimates of the Fisher information that can be biased when the Monte-Carlo noise is non-negligible. Most problematic is noise in the derivatives as this leads to an overestimation of the available constraining power, given by the inverse Fisher information. In this work we find another simple estimate that is oppositely biased and produces an underestimate of the constraining power. This estimator can either be used to give approximate bounds on the parameter constraints or can be combined with the standard estimator to give improved, approximately unbiased estimates. Both the alternative and the combined estimators are asymptotically unbiased so can be also used as a convergence check of the standard approach.  We discuss potential limitations of these estimators and provide methods to assess their reliability. These methods accelerate the convergence of Fisher forecasts, as unbiased estimates can be achieved with fewer Monte Carlo samples, and so can be used to reduce the simulated data set size by several orders of magnitude.
\end{abstract}
\noindent%
\section{Introduction}\label{sec:intro}
A fundamental question across scientific fields is ``How much information can I learn from this experiment?" This arises in diverse situations from designing new experiments to assessing the statistical power of a set of summary statistics to considering the detectability of a new theoretical model. For many of these applications a point estimate of the expected error on a set of parameters is sufficient, and therefore it is desirable to avoid performing a complete statistical inference on mock data -- a process that is frequently complex and time consuming. 

Fisher forecasts are used ubiquitously for this purpose. As a prominent example, many contributions to the recent decadal reviews advising federal agencies on how to plan their use of funds (such as the Astro2020 Decadal Survey on Astronomy and Astrophysics \footnote{https://baas.aas.org/astro2020-science}  and the ``Snowmass" particle physics process\footnote{https://snowmass21.org/}), include Fisher forecasts quantifying the scientific information gain of various experimental or observational projects. 

The underlying principle behind Fisher forecasts is the Cram\'er-Rao bound \citep{Rao_1945,cramer_1946}, which states that the minimum variance of unbiased estimators of the parameters, $\hat{\theta}_i$, is given by
\begin{align}\label{eq:cramerRao}
\mathrm{Var}[\hat{\theta}_i] \geq F^{-1}_{ii},
\end{align}
where $F_{ij}$ is the Fisher Information \citep{Fisher_1922}, defined as
\begin{align}\label{eq:fisherInf}
 F_{ij} = \left< \frac{\partial\log\mathcal{L}(\mathbf{X}|\mathbf{\theta})}{\partial \theta_i} \frac{\partial\log\mathcal{L}(\mathbf{X}|\mathbf{\theta})}{\partial \theta_j} \right>,
\end{align}
and $\mathcal{L}(X|\mathbf{\theta})$ is the likelihood. Fisher forecasts can thus be used to provide fast, point estimates of the constraining power of an experiment.

Beyond its role in computing information bounds, the Fisher information matrix plays a fundamental role in asymptotic statistics \citep{vaart_1998}, and supplies the metric tensor upon which the field of information geometry is built \citep{https://doi.org/10.1111/insr.12464}. 

For many problems analytically computing the Fisher information is intractable and thus it is often estimated using a finite number of simulations of the data. For the often used Gaussian likelihood, the standard formula involves estimating the derivative of the means and the covariance by Monte Carlo. A straightforward evaluation of these quantities gives an estimator that over-predicts the available information. We present an alternative method that provides conservative estimates of the information content. Our new estimator is also biased but, in certain regimes, it underestimates the available information. We show that this estimator is typically biased low by the same amount that the standard estimate is biased high and thus the two estimators can be trivially combined to give unbiased estimates of the Fisher information.

This paper is structured as follows: in \cref{sec:estimators} we review the standard estimator before introducing our new estimators -- the compressed estimator and the combined estimator. To build intuition, in \cref{sec:GaussianExample} we explore the theoretical properties of these estimators when applied to the commonly used Gaussian distribution. In \cref{sec:numerical_example} we demonstrate the advantages of this method on two toy models -- mock experiments with data from a Gaussian distribution and a Poisson distribution. In \cref{sec:est_optimizations} we discuss the practicalities of these estimators and how to test their accuracy. Finally in \cref{sec:test_case} we apply our method to a research example before presenting our conclusions in \cref{sec:conclusions}. Throughout this work we use Einstein summation notation, where repeated indices are summed over.

\section{Three estimators of the Fisher Information}
\label{sec:estimators}
In this Section we first review the standard method for estimating the Fisher information and demonstrate why it is biased high. We then present our alternative estimator, the compressed estimator, and demonstrate why it is biased low. Finally we show how a combined estimator can be trivially formed whose finite-sample bias is greatly reduced.
\subsection{The standard estimator}
\label{sec:standardEst}
To compute the Fisher information, \cref{eq:fisherInf}, we require a likelihood. In this work we consider observables that are part of the exponential family \citep{pitman_1936,Darmois_1935,Koopman_1936}. The exponential family contains many likelihoods used in data analyses, including the Gaussian, Poisson, and Gamma distributions. The likelihood for these observables is can be written in a generic form as
\begin{align}
\mathcal{L}(\mathbf{d}|\mathbf{\theta}) = h(\mathbf{d}) \exp\Big[ \mathbf{\eta}(\mathbf{\theta}) \cdot \mathbf{T}\left(\mathbf{d}\right)-A\left(\mathbf{\eta}(\mathbf{\theta}) \right) \Big],
\end{align}
where $\eta$ is known as the natural parameter, $A$ is the log-partition function and $\mathbf{T}\left(\mathbf{d}\right)$ is the set of sufficient statistics for the distribution. For a given distribution, there are expressions for these parameters in terms of the data and the distribution parameters (e.g., the mean and covariance for a Gaussian).

For this family of estimators the Fisher information is 
\begin{align}
F_{ij} = \frac{\partial \eta_a}{\partial \theta_i}\frac{\partial \eta_b}{\partial \theta_j}\frac{\partial^2 A}{\partial \eta_a \partial \eta_b}= \frac{\partial \eta_a}{\partial \theta_i}\frac{\partial \eta_b}{\partial \theta_j}\mathrm{Cov}\left[T_a,T_b\right].
\end{align}

In the case considered here the covariance of the sufficient statistics and the derivatives of the natural parameters need to be estimated by Monte Carlo. In the standard Fisher estimation approach, the Fisher information is estimated as
\begin{align}
\hat{F}^\mathrm{stnd}_{ij} = \hat{ \eta}_{a,i} \hat{ \eta}_{b,j}\mathrm{Cov}\left[\hat{T}_a,\hat{T}_b\right]
\end{align}
where the covariance matrix and derivatives are estimated from Monte Carlo realizations, hereafter simulations, and for notational simplicity we write derivatives, $\partial f_a/\partial \theta_i$, as $f_{a,i}$. 
In this case we can straightforwardly see that the Fisher information will be biased high
\begin{align}\label{eq:FisherExponential}
\langle\hat{F}^\mathrm{stnd}_{ij}\rangle = F_{ij} + \mathrm{Cov}\left[\hat{ \eta}_{a,i}, \hat{ \eta}_{b,j}\right]\mathrm{Cov}\left[{T}_a,{T}_b\right].
\end{align}
Note that we assume that the simulations for the covariance and the derivatives are independent.

As this bias is always positive, forecast constraints -- given by the inverse -- are always biased low! Given the ingredients for our Monte Carlo Fisher estimate, these bias terms can be estimated and subtracted to obtain an unbiased estimate of the Fisher information. However,   the unbiased estimate obtained from subtracting the biases is not useful as the resulting matrix is generally not invertible and thus we cannot use the estimate to infer parameter constraints.

\subsection{The compressed estimator}
\label{sec:compressedEst}
Motivated by this issue we consider an alternative, a compressed Fisher forecast. 
Consider the compressed statistic
\begin{align}\label{eq:compression_general}
t_i =\left. \frac{\partial \ln\mathcal{L}(\mathbf{X}|\theta)}{\partial \theta_i}\right\vert_{\theta_*},
\end{align}
where $\theta_*$ denotes the parameter values where the score is evaluated \citep{Tegmark_1997,Heavens_2000,Alsing_2018}. This compression is optimal at $\theta_*$, thus
\begin{align}
F^\mathrm{comp.}_{ij} = \left< \frac{\partial\log\mathcal{L}(\mathbf{t}|\mathbf{\theta})}{\partial \theta_i} \frac{\partial\log\mathcal{L}(\mathbf{t}|\mathbf{\theta})}{\partial \theta_j} \right> = F_{ij},
\end{align}
i.e. the compression loses no information. As is discussed in \citet{Alsing_2018}, the compressed statistics in \cref{eq:compression_general} are linearly related to quasi maximum-likelihood estimators and as such are asymptotically normally distributed. Thus the Fisher information for these statistics can be approximated as
\begin{align}\label{eq:fisherCompressedGaussApprox}
F^\mathrm{comp.}_{ij} = \mu^t_{a,i} {\Sigma}^{-1}_{ab} \mu^t_{b,j} 
\end{align}
where $\mu^t_i=\langle t_i\rangle$ and $ {\Sigma}_{ab}= \langle (t-\mu^t)_a(t-\mu^t)_b \rangle$. 

As in the standard estimator, these components will often need to be estimated with simulations as
\begin{align}
\hat{F}^\mathrm{comp.}_{ij} = \hat{\mu}^t_{a,i} {\hat{\Sigma}}^{-1}_{ab} \hat{\mu}^t_{b,j} .
\end{align}

The key proposal in this work is, when performing Monte Carlo estimates of the Fisher information, to estimate the compressed Fisher information -- using the Gaussian approximation \cref{eq:fisherCompressedGaussApprox} when needed. This is of most utility when further used to compute the covariance of parameters. The intuition is that the compressed statistics combine all the data points in the noisy vectors and extract the information relevant for the parameter constraints. The noise in the derivatives and covariance for the compressed statistics will thus be averaged down. This can be exploited to obtain a lower noise estimate for the Fisher information. This can be seen by comparing the expectation of the standard Fisher, \cref{eq:FisherExponential}, with the corresponding expectation of the compressed Fisher

\begin{align}\label{eq:expectationFisherCompressed}
\langle \hat{F}^\mathrm{comp.}_{ij}\rangle &={F}_{ij}+ {{\Sigma}}^{-1}_{ab} \mathrm{Cov}\left[{\mu}^t_{a,i},{\mu^t}_{b,j}\right]
\end{align}
- both Fisher estimates are biased high by the traces of covariance matrices. However for the standard case the dimension of the matrix is the data dimension, whilst for the compressed case it is the dimension of the parameters. This leads to a large reduction in the bias. We explore how this works in detail for a Gaussian likelihood in \cref{sec:intutitionGaussian}. 

A challenge of this approach is that the optimal compression, \cref{eq:compression_general}, requires access to the same quantities as the Fisher information -- the compression is given by the score and the Fisher information the variance of the score! For example, compression with a Gaussian likelihood (with parameter independent covariance matrix) is given by
\begin{align}\label{eq:compression}
t_i = \mu_{a,i} C^{-1}_{ab}\left({d_b- \mu_b}\right)
\end{align}
which requires derivatives of the mean and covariance matrix and the data covariance matrix -- if we had perfect knowledge of these we could perform the Fisher analysis. 

This circular dependency can be resolved by using an approximate compression scheme. When using approximate compression methods, the compressed Fisher information will no longer be optimal and thus the resulting estimate of the Fisher information (parameter errors) will be an underestimate (overestimate) \citep[see e.g.,][]{Lehmann_2006}. This can be useful as it provides an estimate of the minimum information that can be learnt from the data set. 

The approximate compression method we consider is simply to replace the unknown components in the optimal compression, \cref{eq:compression_general}, with Monte Carlo estimates. This has two nice features: first it asymptotes to an optimal compression with increasing size of the Monte Carlo. Second it requires the same types of Monte Carlo as are required for the Fisher estimate -- thus no additional data products need to be simulated. To avoid biases the Monte Carlo simulations used for the compression should be different from those used in estimating the compressed Fisher forecasts. Thus we propose a two step process: first take a portion of the simulations and use those simulations to compute the components required for the compression. Second use the remainder of the simulations to compute the compressed Fisher information. See \cref{sec:divisionOfSims} for a discussion of how to divide the simulations.

Note that this compression scheme is a random compression scheme as the compression function will depend on the realizations of simulations used to estimate it. This will add additional noise to our compressed Fisher estimate as the suboptimality of the compression is slightly different for each realization. We discuss this further in \cref{sec:resampling}. 

\subsection{Combined Estimators}
There are two interesting aspects of the compressed Fisher forecast. First, the derivative of the mean of the compressed statistic, ${\mu}^t_{a,i}$, is an estimator of the Fisher information information! This trivially follows from our choice of compression
\begin{align}\label{eq:meanCompressed}
 {\mu}^t_{a,i}= \frac{\partial t_a}{\partial \theta_i} = \left.\frac{\partial^2 \ln \mathcal{L}}{\partial \theta_i \partial\theta_a}\right\vert_{\theta_*} = \hat{F}^\mathrm{deriv}_{a,i}
\end{align}
and from the relation
\begin{align}\label{eq:FihserCovMatCompressed}
\left< \frac{\partial^2\log\mathcal{L}(\mathbf{X}|\mathbf{\theta})}{\partial \theta_I\theta_J} \right> = \left< \frac{\partial\log\mathcal{L}(\mathbf{X}|\mathbf{\theta})}{\partial \theta_I} \frac{\partial\log\mathcal{L}(\mathbf{X}|\mathbf{\theta})}{\partial \theta_J} \right> = F_{IJ},
\end{align}
which holds when the log-likelihood is twice differentiable and subject to regularization conditions \citep{Fisher_1922}. If computed by a Monte Carlo estimate, this is an unbiased estimate of the Fisher information, but generally not invertible. 

Second the variance of the compressed statistic is also an estimate of the Fisher information, which is straightforwardly seen as
\begin{align}\label{eq:varCompressed}
\Sigma_{ij} = \langle t_i t_j \rangle = \langle \frac{\partial \ln\mathcal{L}}{\partial \theta_i} \frac{\partial \ln\mathcal{L}}{\partial \theta_j} \rangle= \hat{F}^\mathrm{stnd}_{ij},
\end{align}
where we have used that $\langle t_i\rangle =0$. Estimating the variance of the compressed statistic is exactly equivalent to the standard Fisher information estimator.

These pieces offer a complementary view of the compressed Fisher estimator: the compressed Fisher estimate can be thought of as effectively the square of an unbiased and non-invertible Fisher estimate (the derivative terms) normalized by a biased and invertible estimate (the variance term), i.e.
\begin{align}
\hat{F}^\mathrm{comp.}_{IJ} = \hat{F}^\mathrm{deriv}_{Ia}\widehat{ \left( \left.\hat{F}^\mathrm{stnd}_{ab}\right.^{-1}\right)} \hat{F}^\mathrm{deriv}_{bJ}
\end{align} 
with expectation
\begin{align}\label{eq:compressed_fish_v2}
  \langle \hat{F}^\mathrm{comp.}_{IJ} \rangle = F_{Ia} \left(\langle \hat{F}^\mathrm{stnd}_{ab}\rangle\right)^{-1}  F_{bJ} + \left(\langle \hat{F}^\mathrm{stnd}_{ab}\rangle\right)^{-1} \mathrm{Cov}\left[\hat{F}^\mathrm{deriv}_{Ia},\hat{F}^\mathrm{deriv}_{bJ}\right].
\end{align}
As discussed above, and demonstrated explicitly in \cref{sec:intutitionGaussian}, the additive noise bias is smaller than the noise bias in the standard estimator. Thus this bias can be made negligible with a small number of Monte-Carlo realizations. 

When in the regime where the noise is subdominant, we can see the low bias  of the compressed estimate, which arises from the suboptimal and lossy compression, is exactly introduced by the inverse of the standard Fisher estimate \cref{eq:varCompressed}. Motivated by this we also propose a combined estimator that is the geometric mean of the two estimators, i.e.
\begin{align}\label{eq:combinedEstimate}
F^\mathrm{Combined}_{IJ} = G(F^\mathrm{Standard}_{IJ},F^\mathrm{comp.}_{IJ})
\end{align}
where $G(A,B)$ is the geometric mean of matrix $A$ and $B$, defined as \citep{bhatia07}
\begin{align}
G(A,B) = A^{\frac{1}{2}} \left(A^{-\frac{1}{2}}B A^{-\frac{1}{2}} \right)^{\frac{1}{2}} A^{\frac{1}{2}}. 
\end{align}

The geometric mean estimator uses the high bias of one estimator to cancel the low bias of the other estimator, resulting in an unbiased estimate of Fisher information.

One point to note is that the degree of suboptimality is set by the number of simulations used in the compression step. Thus to obtain an unbiased combined estimator, we need to combine the compressed estimator with the standard estimator computed using the same number of simulations as in the compression step.

\section{Gaussian Fisher Forecast}\label{sec:GaussianExample}
As an example of how this works in practice lets consider the case of a Gaussian likelihood.
\subsection{Standard Fisher Estimate}
For a Gaussian likelihood the Fisher information is
\begin{align} \label{eq:true_fisher}
F_{ij} = \mu_{a,i} C^{-1}_{ab} \mu_{b,j} + \frac{1}{2}\left[C^{-1}_{ab}C_{bc,i}C^{-1}_{cd}C_{ca,j} \right]
\end{align}
where $\mu_a$ is the observable mean, $C_{ab}$ is the data covariance matrix -- both $\mu$ and $C$ are functions of the parameters of interest.

To estimate the Fisher information for the case of a Gaussian likelihood we require three components, the covariance matrix of the observables, the derivative of the mean of the observable and the covariance matrix with respect to the parameters of interest. A naive estimate of the Fisher information using Monte Carlo realizations would be
\begin{align} \label{eq:standard_fisher}
\hat{F}_{ij} = \hat{\mu}_{a,i} \tilde{C}^{-1}_{ab} \hat{\mu}_{b,j} + \frac{1}{2}\left[\tilde{C}^{-1}_{ab}\hat{C}_{bc,i}\tilde{C}^{-1}_{cd}\hat{C}_{da,j} \right].
\end{align}
Note we use 
\begin{align}
\tilde{C}^{-1}_{ab} = \frac{n_s-d-2}{n_s-1} \hat{C}^{-1}_{ab},
\end{align}
where $ \hat{C}$ is the sample covariance matrix, to avoid biases in the application of the inverse sample covariance matrix \citep{Anderson_2003,Hartlap_2007}.
We can see immediately that the Fisher information will biased high as
\begin{align}\label{eq:expectionStandardFisherBiased}
\langle \hat{F}_{ij}\rangle &= F_{ij}\nonumber \\ 
&+ {C}^{-1}_{ab} \mathrm{Cov}\left[\delta{\mu}_{a,i},{\delta\mu}_{b,j}\right]+\frac{1}{2} C^{-1}_{ab}C^{-1}_{cd} \mathrm{Cov}\left[{\delta C}_{bc,i},{\delta C}_{da,j}\right] \nonumber \\ 
&+ \frac{1}{2} \mathrm{Cov}\left[\tilde{C}^{-1}_{ab},\tilde{C}^{-1}_{cd} \right]\Big[{C}_{bc,i}{C}_{da,j} + \mathrm{Cov}\left[{\delta C}_{bc,i},{\delta C}_{da,j}\right] \Big].
\end{align}
There are three sources: the variance of the derivatives of the mean, the variance of the derivatives of the covariance matrix and the variance of the precision matrix.For the oft-encountered case where the covariance matrix is independent of the parameters of interest, this simplifies to
\begin{align}\label{eq:expectionStandardFisherBiased_noCovMat}
\langle \hat{F}_{ij}\rangle &= F_{ij}+ {C}^{-1}_{ab} \mathrm{Cov}\left[\delta{\mu}_{a,i},{\delta\mu}_{b,j}\right].
\end{align}
\subsection{The compressed Gaussian Fisher Information}
The optimal compression for the Gaussian case is given in \cref{eq:compression} and cannot be evaluated without knowledge of the quantities required for the Fisher forecast. 
Hence we use a subset of the simulations, labeled by superscript $\alpha$ to compute the suboptimal compression
\begin{align}
t_i &= \mu^{\alpha}_{a,i} {C^\alpha}^{-1}_{ab}\left({d_b- \mu^{\alpha}_b}\right)\nonumber \\ 
&+\frac{1}{2} (d_a-\mu^{\alpha}_a){C^\alpha}^{-1}_{ab}{C^\alpha}_{bc,i}{C^\alpha}^{-1}_{cd}(d_d-\mu^{\alpha}_d) -\frac{1}{2} {C^\alpha}^{-1}_{ab} {C^\alpha}_{ba,i}
\end{align}
We then use the remainder of the simulations, labelled with superscript $\beta$, to compute the compressed Fisher components i.e.
\begin{align}
\hat{\mu}^t_{i,I} = \mu^{\alpha}_{a,i} {C^\alpha}^{-1}_{ab}\mu^\beta_{b,I}+\frac{1}{2} C^\beta_{ad,I}{C^\alpha}^{-1}_{ab}{C^\alpha}_{bc,i}{C^\alpha}^{-1}_{cd},
\end{align}
 and 
\begin{align}
 \hat{\Sigma}_{ij}&= \mu^{\alpha}_{a,i} {C^\alpha}^{-1}_{ab}C^\beta_{bc,I}{C^\alpha}^{-1}_{cd}\mu^\alpha_{d,j}\nonumber \\ 
&+\frac{1}{2} {C^\alpha}^{-1}_{ab}{C^\alpha}_{bc,i}{C^\alpha}^{-1}_{cd} {C^\alpha}^{-1}_{AB}{C^\alpha}_{BC,j}{C^\alpha}^{-1}_{CD}\left[C^\beta_{aA}C^\beta_{dD} \right],
\end{align}
where $n_\beta$ is the number of simulations in the set $\beta$. Note the covariance matrix can equally be written as
\begin{align}\label{eq:compressedVarianceAlt}
 \hat{\Sigma}_{ij} = \mu^{\alpha}_{a,i} {C^\alpha}^{-1}_{ab}\mu^{\alpha}_{b,j} + \frac{1}{2}\left[ {C^\alpha}^{-1}_{ab} {C^\alpha}_{bc,i} {C^\alpha}^{-1}_{cd} {C^\alpha}_{da,j} \right],
\end{align}
which is just the uncompressed Fisher information. We found this form provided more optimal Fisher estimates as it is the lower noise estimate.

For the case where the covariance matrix is independent of the parameters these equations simplify to
\begin{align}
\hat{\mu}^t_{i,I} = \mu^{\alpha}_{a,i} {C^\alpha}^{-1}_{ab}\mu^\beta_{b,I},
\end{align}
and
\begin{align}\label{eq:compressedVarianceAlt_noCovMat}
 \hat{\Sigma}_{ij} = \mu^{\alpha}_{a,i} {C^\alpha}^{-1}_{ab}\mu^{\alpha}_{b,j}.
\end{align}
\subsection{Building intuition}\label{sec:intutitionGaussian}
To help build intuition into why this may be helpful consider a simplified case where 
\begin{align}
 \mathrm{Cov}\left[\delta{\mu}_{a,i},{\delta\mu}_{b,j}\right] = \frac{1}{N} \zeta_{ij}C_{ab}
\end{align}
\begin{align}
C_{ab,i} = 0
\end{align}
where $N$ is the number of simulations used to estimate the derivatives and $\zeta_{ij}$ is a matrix relating the scale of the noise in the derivatives to the noise in the data vector. For this case we assume the noise on the covariance matrix is subdominant. Whilst this is a very simple case it highlights some key features and can arise when the noise on the data is weakly dependent on the parameters. An example where this is approximately the case is
when the derivatives are estimated by finite differencing, i.e.
\begin{align}
\hat{\mu}_{a,i} = \frac{1}{2 \delta \theta_i} \Big(\hat{d}_a |_{\theta=\theta_*+\delta \theta_i}- \hat{d}_a |_{\theta=\theta_*-\delta \theta_i} \Big).
\end{align}
Note that the assumption of $C_{ab,i} = 0$ is frequently used in physics and is motivated by the work of \citet{Carron_2013}.

Under these assumptions we see that the bias to the uncompressed Fisher information is
\begin{align}
\langle F^\mathrm{stnd}_{ij} \rangle = F_{ij} +\frac{d}{N}\zeta_{ij}.
\end{align}
and the compressed Fisher information is 
\begin{align}
\langle F^\mathrm{comp.}_{ij} \rangle = F^\mathrm{comp.}_{ij} +\frac{p}{N}\zeta_{ij}.
\end{align}
A comparison of these two equations demonstrates the salient features of our approach. First, the additive bias to the compressed Fisher is reduced by the ratio of the dimension of the data vector, $d$, to the number of parameters $p$. Second,  as
\begin{align}
    F^\mathrm{comp.}_{ij} \approx F^{iI} \langle F^\mathrm{std}_{IJ}\rangle^{-1} F^{Jj},
\end{align}
the compressed Fisher information is biased low by approximately the same amount the standard Fisher is biased high.

Thus when the additive bias to the compressed estimator is negligible, the geometric mean estimator --\cref{eq:combinedEstimate}-- will provide an approximately unbiased constraint. This example demonstrates that the additive bias to the compressed Fisher information is expected to be significantly smaller ( by a factor $p/d$) than the additive bias to the standard Fisher information. Therefore, we expect many cases where the standard Fisher information is biased, but the geometric estimator is unbiased.

\section{Numerical Examples} \label{sec:numerical_example}
To demonstrate our method we apply it to two examples; the first is for data from a Gaussian distribution and the second is for a Poisson distribution.
\subsection{Gaussian Likelihood}
For the first case, consider data drawn from a normal distribution, $d(x) \sim \mathcal{N}(\mu(x,\theta),C(k,k'))$, where the mean,
\begin{align}
\mu(x,\theta) = \alpha +\beta x+\gamma x^\frac{1}{2},
\end{align}
and the covariance matrix,
 \begin{align}
 C(x,x') = \delta(x-x') 2 \mu(x,\theta)^2,
 \end{align}
  are functions of three parameters ($\alpha$, $\beta$ and $\gamma$).
  
We consider the data vector from points sampled at $100$ spatial locations, $x$, logarithmically from $10^{-4}$ to $1$. We consider the Fisher information about the fiducial parameters $\theta_*=(1,1,1)$. This setup was chosen semi-arbitrarily as it exhibits non-trivial degeneracies often found in data analyses.
 
 To perform the standard Fisher analysis we need to estimate the derivative of the mean with respect to the parameters and the covariance matrix. 
 We estimate the derivatives with finite difference as
\begin{align}\label{eq:finiteDifDerivative}
\hat{\mu}_{a,i} = \frac{1}{2 \delta \theta_i} \Big(\hat{d}_a |_{\theta=\theta_*+\delta \theta_i}- \hat{d}_a |_{\theta=\theta_*-\delta \theta_i} \Big).
\end{align}
and
\begin{align}\label{eq:finiteDifDerivativeCovMat}
\hat{C}_{ab,i} = \frac{1}{2 \delta \theta_i} \Big(\hat{C}_{ab} |_{\theta=\theta_*+\delta \theta_i}- \hat{C}_{ab} |_{\theta=\theta_*-\delta \theta_i} \Big)
\end{align}
where $\delta \theta_i = 0.1$ and $\hat{C}$ is the covariance matrix estimated from simulations. To demonstrate a realistic case we employ a commonly used variance cancellation method: the seeds used by the random number generators for the Monte Carlos simulations at $\theta+\delta \theta$ and $\theta-\delta \theta$ are the same. Matching the seeds cancels most of the noise in the derivatives and is commonly used in simulation-based Fisher forecasts \citep[see e.g.,][for an example]{villaescusa-Navarro_2020}. The covariance matrix is estimated from an ensemble of simulations with $N_{\mathrm{cov\,.mat.}} = 5000$. 

For the compressed forecast we split the derivative simulations into two parts: $90\%$ are used for the compression and the remainder half for the derivatives of the compressed statistics. We use \cref{eq:compressedVarianceAlt} to compute the covariance matrix of the compressed statistics.

\begin{figure*}
 \centering
 \subfloat[Fisher Information ]{\label{fig:fisher_3param}%
 \includegraphics[width=0.45\textwidth]{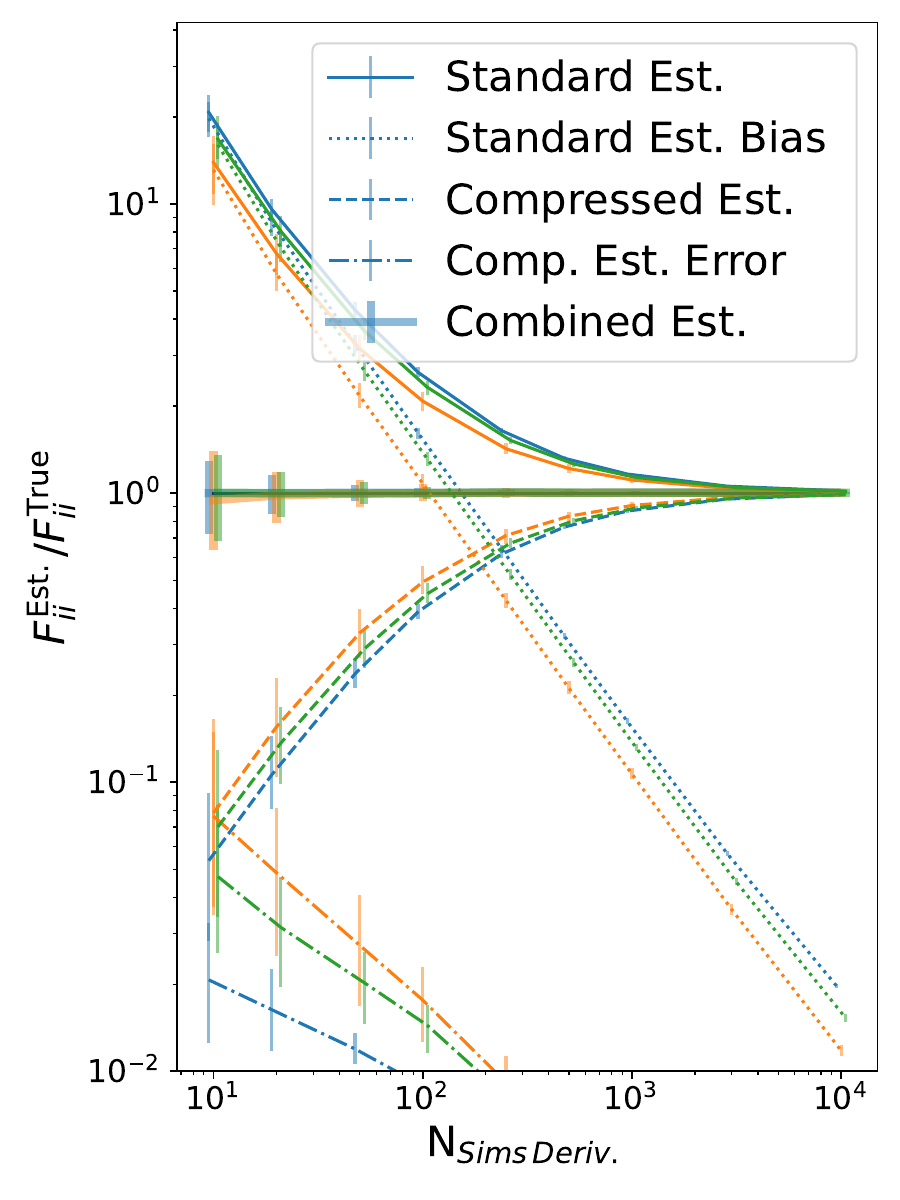}%
}
 \subfloat[Parameter Constraints]{\label{fig:constraint_3param}%
 \includegraphics[width=0.45\textwidth]{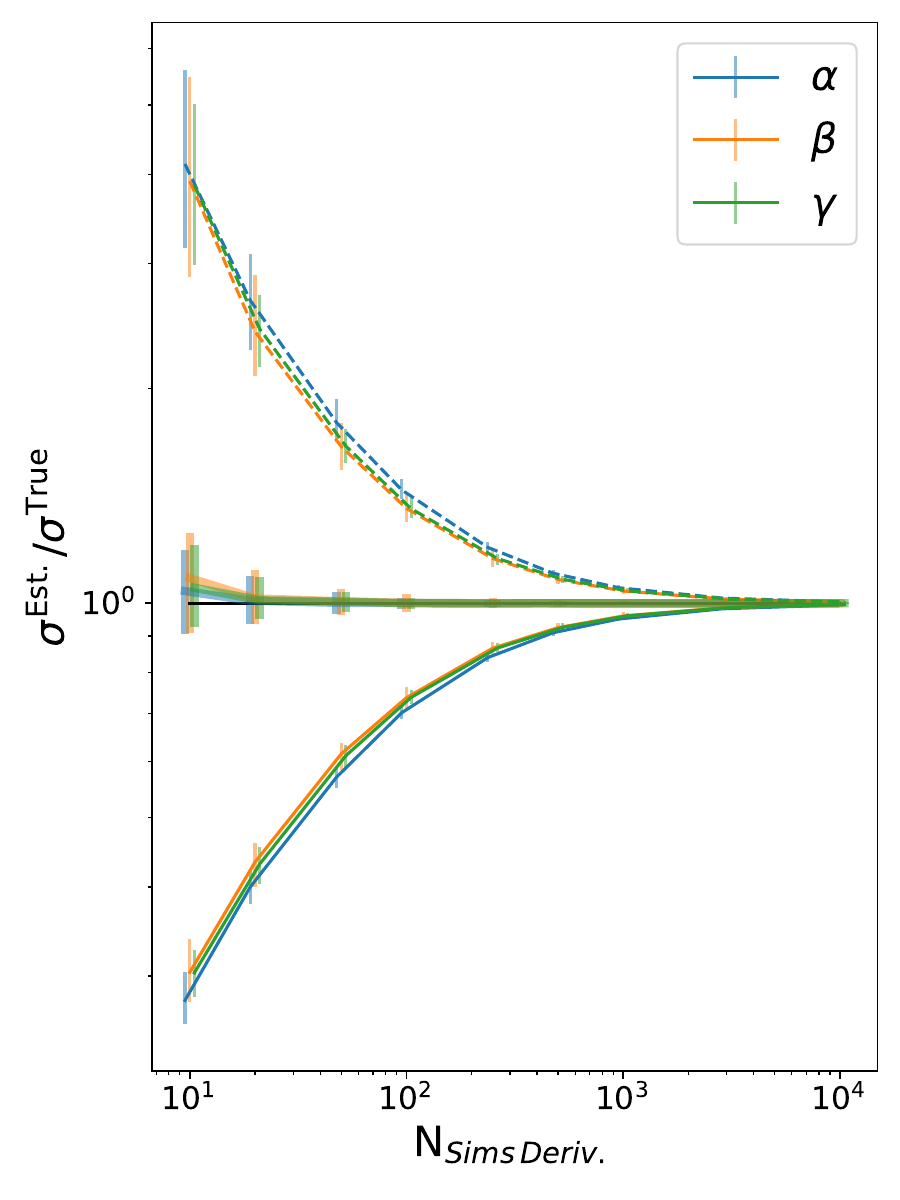}%
} 
 \caption{In \cref{fig:fisher_3param} the ratio of the estimated Fisher information to the truth for the Gaussian toy model is plotted as a function of the number of simulations used to estimate the derivatives. We compare the standard estimator (thin lines), the compressed Fisher information (dashed lines) and the combined estimator (thick lines). In dotted lines we estimate the bias to the standard estimator, \cref{eq:expectionStandardFisherBiased}, and in dot-dashed to the compressed estimator, \cref{eq:expectationFisherCompressed}. The standard and compressed estimates are both biased estimates of the Fisher information. However note that the additive noise biases to each estimator are two orders of magnitude smaller for the compressed estimator. The combined estimator is effectively unbiased.  In \cref{fig:constraint_3param} we plot the ratio of the estimated constraints (thin lines for the standard estimator, dashed for the compressed estimate and thick lines for the combined estimator ) to the true Fisher constraint. As expected the standard estimator is biased low, the compressed is biased high and the combined provides an accurate estimate. The errorbars denote the $1\sigma$ credible interval about the median. \label{fig:GaussianPlots} }
\end{figure*}

In \cref{fig:fisher_3param} we compare the standard and compressed Fisher estimates to the truth. Firstly we note that, as expected the standard Fisher estimate is biased high. Second we note that the approximation to the bias, \cref{eq:expectionStandardFisherBiased} evaluated using Monte Carlo products, accurately estimates the bias. Thus it is straightforward to estimate the unbiased Fisher information. However subtracting this bias, generically, does not lead to an invertible matrix and so cannot be used to estimate the parameter covariance matrix. Next we see that the compressed Fisher is biased low (as expected). Using \cref{eq:expectationFisherCompressed} we can also estimate the noise bias on the compressed estimator finding that it also accurately matches the observed bias.  Finally we also plot the combined estimator, \cref{eq:combinedEstimate}, we see that across the entire range this gives an unbiased estimate of the Fisher information.
 
In \cref{fig:constraint_3param} we convert the Fisher information estimates into estimates of the forecast parameter variances. We see that when a small number of simulations are used the standard estimator underestimates the parameter variance. On the other-hand the compressed Fisher estimator overestimates the error, due to the suboptimal compression. As more simulations are included, the biases to the standard estimator become more and more subdominant and the estimate tends to the true error from below. Adding more simulations to the compressed estimator improves the compression and thus this estimator tends to the error forecast from the true Fisher information from above. As with the Fisher information, unbiased forecast constraints can be obtained with the combined estimator!

\subsection{Poisson Model}\label{sec:poissonModel}
As a second common example consider the case of data from a Poisson distribution: $\mathbf{d}(x) \sim \mathrm{Pois}(\lambda(x))$, where the rate parameter, $\lambda(x)$, has the same form as the mean function for the Gaussian case
\begin{align}
\mu(x,\theta) = \alpha +\beta x+\gamma x^\frac{1}{2}.
\end{align}
It is also function of three parameters ($\alpha$, $\beta$ and $\gamma$). In this case we consider a data vector sampled at $100$ spatial, $x$, points sampled logarithmically from $10^{-4}$ to $1$. We consider the Fisher information about the fiducial parameters $\theta_*=(1,1,1)$. In this case we do not use the `matching' procedure used above and hence require many more simulations.

The standard Fisher information is 
\begin{align}\label{eq:fisher_poisson}
F_{ij} = \sum\limits_{x}\frac{\partial \ln \lambda(x)}{\partial \theta_i} \frac{\partial \ln \lambda(x)}{\partial \theta_j} \mathrm{Var}[d(x)].
\end{align}
and the optimal compression is given by
\begin{align}\label{eq:compression_poisson}
t_i =\sum\limits_{x} \left(d(x)-\lambda(x) \right)\frac{\partial\ln \lambda(x)}{\partial \theta_i}
\end{align}
As before we estimate the mean, the variance and the derivatives using Monte Carlo simulations. The derivatives use the finite differences as in \cref{eq:finiteDifDerivative} with steps $\delta \theta =0.05$ . As in the Gaussian case we use part of the simulations (in this case 50 \%) to compute the quantities for the compression and then, with the remainder of the simulations, we estimate the compressed Fisher information, \cref{eq:fisherCompressedGaussApprox}. 

\begin{figure*}
 \centering
 \subfloat[Fisher Information ]{\label{fig:fisher_3param_poisson}%
 \includegraphics[width=0.45\textwidth]{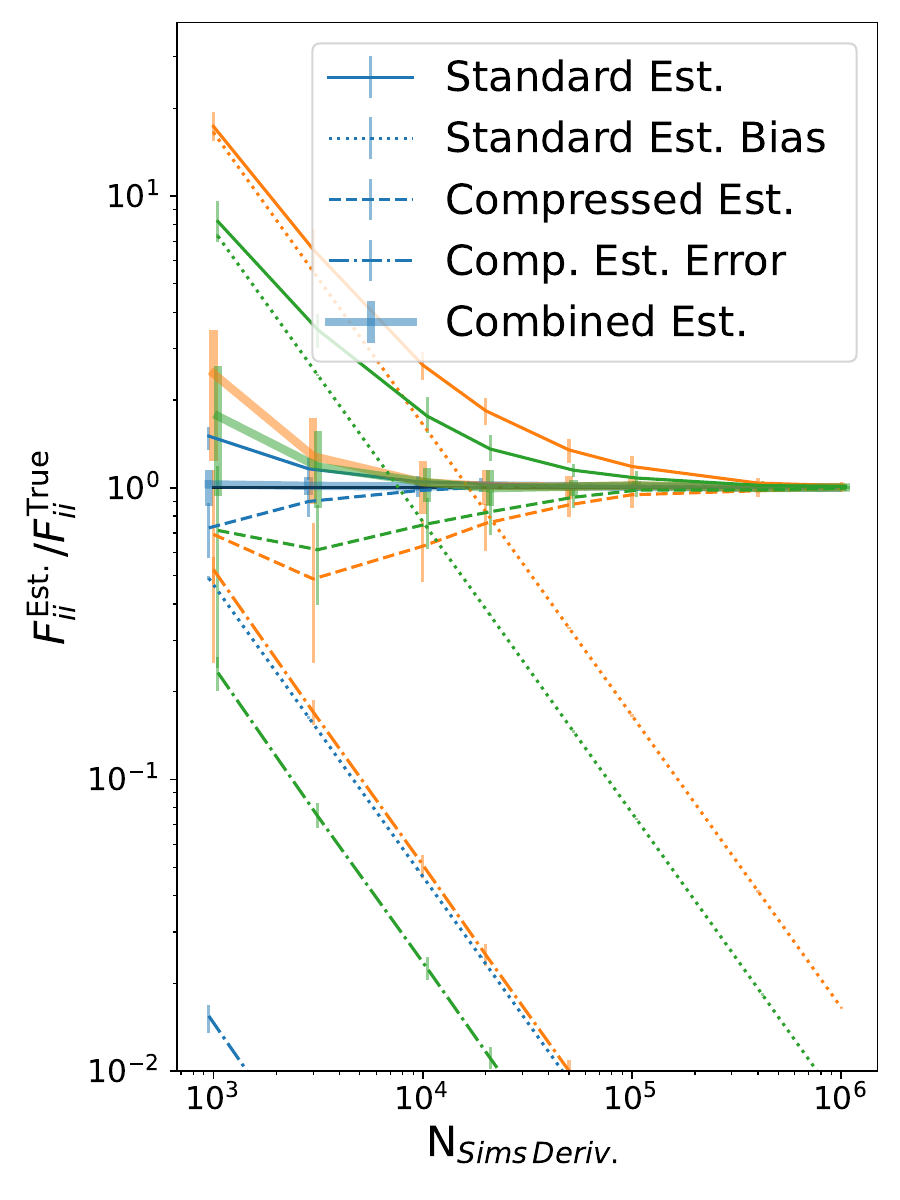}%
}
 \subfloat[Parameter Constraints]{\label{fig:constraint_3param_poisson}%
 \includegraphics[width=0.45\textwidth]{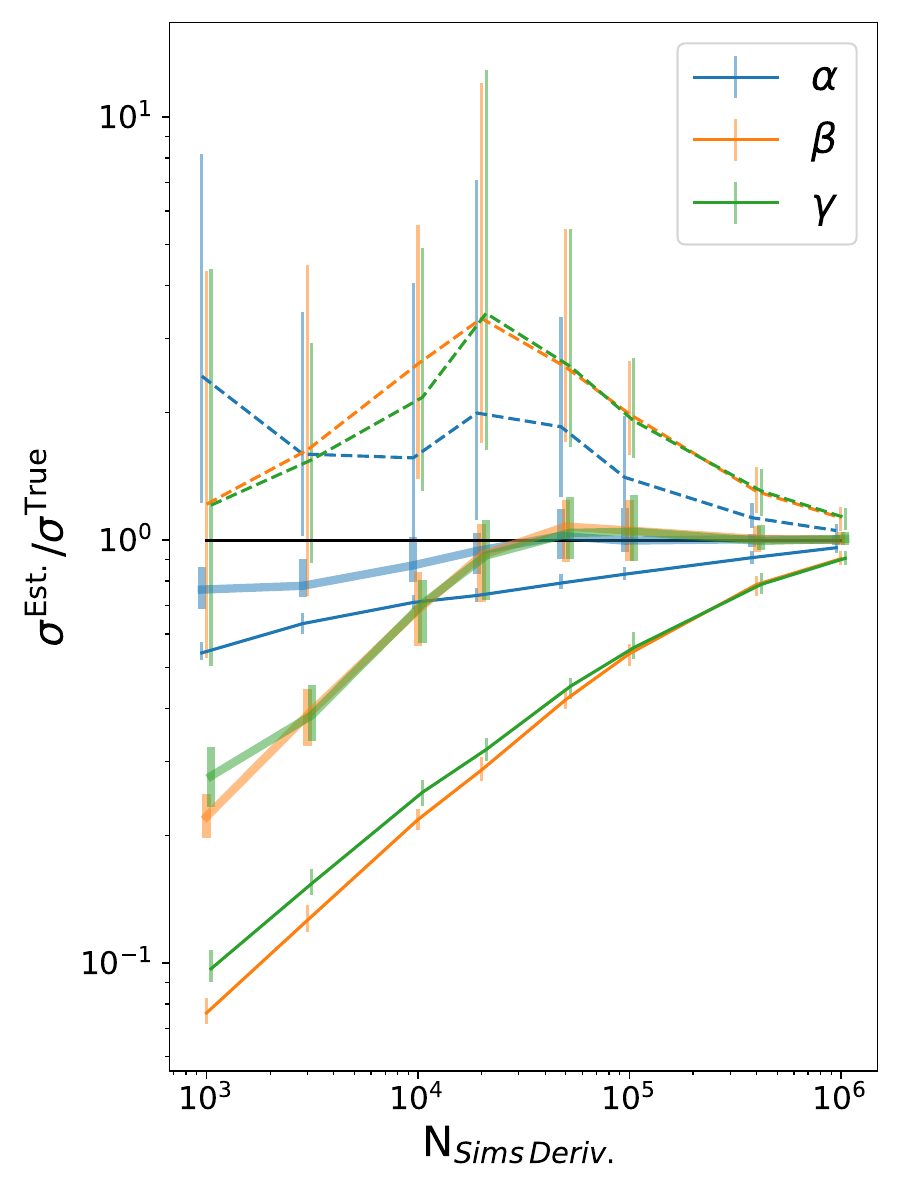}%
} 
 \caption{Plots of the Fisher Information and parameter constraints, as in \cref{fig:GaussianPlots}, but for the case of a Poisson distribution. Note that, unlike the Gaussian case, we do not use the `seed' matching case for the derivatives (see \cref{sec:GaussianExample} for details). This results in significantly larger noise in the derivatives. For low numbers of simulations all the estimators are biased. However the combined estimator strongly accelerates convergence of the  estimates with orders of magnitude fewer simulations than the standard method. \label{fig:PoissonPlots} } 
\end{figure*}

In \cref{fig:fisher_3param_poisson} we compare the two Fisher estimates to the truth. The standard estimator shows qualitatively the same features as the Gaussian example: it is biased high, in a manner that is calculable. The compressed Fisher shows more complex behavior: for small numbers of simulations it is also biased high. This a result of the bias terms in \cref{eq:expectationFisherCompressed} -- note that these biases are still drastically smaller than the uncompressed case. Using \cref{eq:expectationFisherCompressed} we can also estimate the bias on the compressed estimator finding that it also accurately matches the observed bias. 
As more simulations are included this bias becomes subdominant and the situation becomes similar to the Gaussian case. In this case we only consider the combined estimator when this bias is subdominant. In this regime the results are similar to the Gaussian case -- the combined estimator gives unbiased estimates of the Fisher information.

In \cref{fig:constraint_3param_poisson} we convert the Fisher information estimates into estimates of the parameter variances. We see that when a small number of simulations are used both estimates underestimate the parameter variance. As the biases seen in the Fisher information estimates become negligible (see \cref{sec:biasImportance}), the two estimators tend towards the true value, with the compressed estimator providing an over estimate and the standard estimate an underestimate. In this regime the combined estimator can be applied and it greatly accelerates the Monte-Carlo convergence of the Fisher errors.

\section{The practicalities of the combined estimator}\label{sec:est_optimizations}
Given analytical and numerical evidence for the faster convergence of the combined estimator, in this section we discuss three aspects of using this estimator in practice: what fraction of the simulations should be used for the compression step vs (\cref{sec:divisionOfSims}), the value of averaging multiple different splits of the simulations (\cref{sec:resampling}) and how to assess if the noise bias term is subdominant -- and thus the combined estimator is unbiased (\cref{sec:biasImportance}).  
\subsection{How many simulations to use for the compression step?}\label{sec:divisionOfSims}
In principle, there is substantial freedom in choosing how to split the simulations between the compression and the estimation of the derivatives. The combined estimator is nearly unbiased providing the additive noise bias, the second term in \cref{eq:compressed_fish_v2}, is subdominant. This suggests that we want to use the majority of the simulations to estimate the compressed Fisher and a smaller fraction in the compression. However, there is a slight subtlety -- using fewer simulations in the compression also reduces the Fisher information. This reduction in the compressed Fisher information means that despite the reduction in the noise bias, from using more simulations to estimate the compressed information, the relative importance of the noise bias can be similar. For the cases considered here that was generally true -- if the noise bias was important for one choice of the division of simulations then it was not often possible to choose an alternative division to mitigate it. Likewise the reverse was true -- if the noise was subdominant, it generally remained so for reasonable divisions of simulations. The term ``reasonable divisions" means that we did not assign 99\% of the simulations to one task -- in those regimes the noise bias term could become dominant. If one is interested in using just the compressed estimator the situation is slightly different and we found that altering the fractions could slightly reduce its level of bias. 

In summary, when using the recommended, combined estimator the results are fairly insensitive to the division of simulations between constructing the compression and estimating the compressed Fisher information. Thus, it is reasonable to start with 50\% of the simulations for each task and to perform adjustments to fine tune the results.

\subsection{Shuffling the simulations}\label{sec:resampling}
As in evident in \cref{fig:constraint_3param} and \cref{fig:constraint_3param_poisson} the variance of the compressed estimator is significantly larger than the standard case! Fortunately the combined estimator does not inherit the O(1) variance of the compressed estimator. However it still has a significantly larger variance, by up to ~10\%, than the standard Fisher estimate. 

Part of this extra noise arises as the compression is noisy. We found that part of the combined estimators extra variance can be reduced by using the freedom we have in assigning simulations to the compression and Fisher estimation steps. Having performed one estimate of the combined Fisher information, we can obtain a second estimate by choosing a different assignment of the simulations to the compression and Fisher estimation steps. This can be repeated multiple times to obtain a set of many, partially correlated, estimates that can then be averaged. 

\begin{figure}
\centering
 \includegraphics[width=0.45\textwidth]{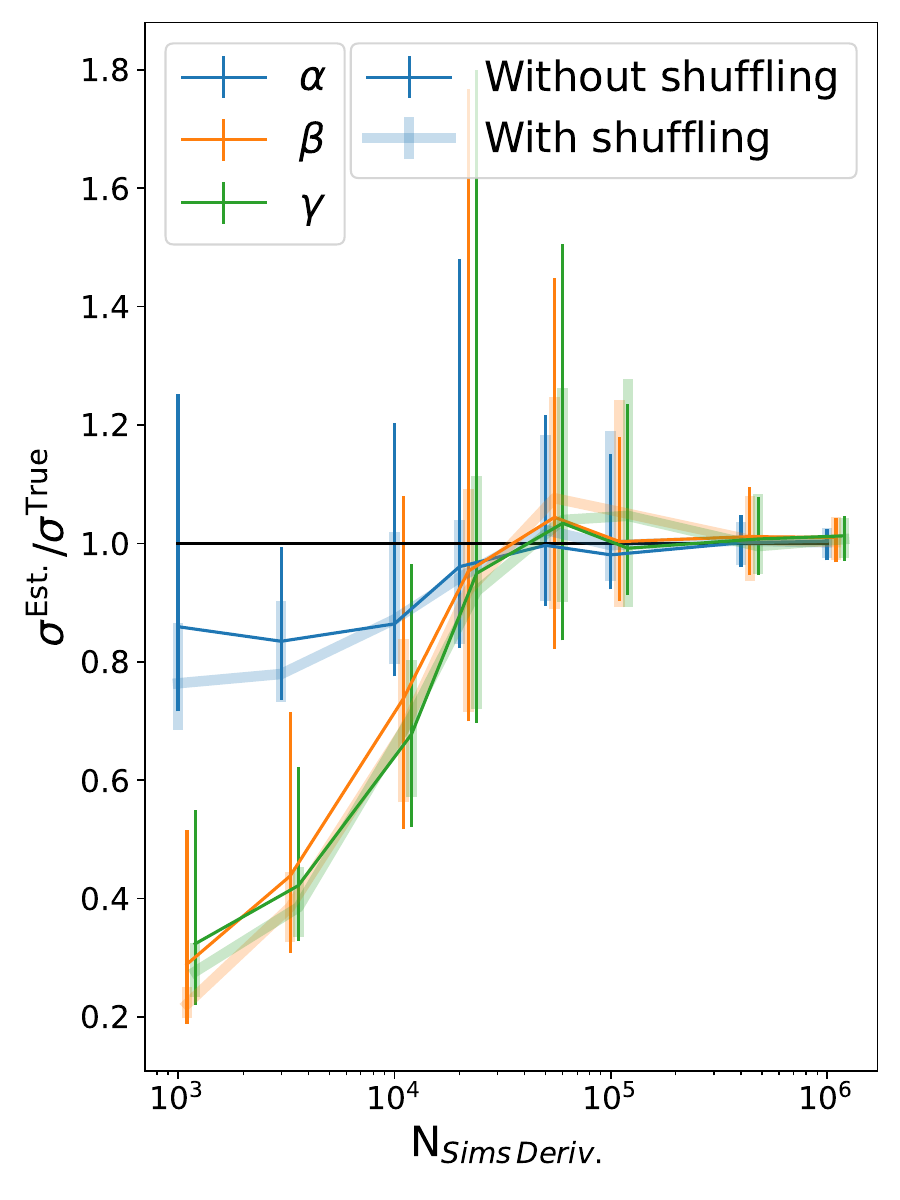}
 \caption{ The compressed and combined estimators exhibit a larger variance than the standard, biased Fisher estimator. By repeatedly shuffling simulations between the compression and Fisher estimation steps, and then averaging we can reduce the variance on these estimators. Here we apply the combined estimator without shuffling, thin lines, and the estimator averaged with 10 shufflings, thick lines, to the Poisson example, \cref{sec:poissonModel}. The estimator without shuffling shows a larger variance when a small number of simulations are used.
 \label{fig:resampling}} 
\end{figure}

The benefits of this approach can be seen in \cref{fig:resampling}. Here we show the combined estimator applied to the Poisson model from \cref{sec:poissonModel}. We show two cases of this model: one using only one division of the simulations and a second using an average of ten shufflings of the simulations. For small numbers of simulations, the shuffling operation can dramatically reduce the estimator variance. On the other hand, for large numbers of simulations the shuffling has no effect on the variance.

\subsection{When can we trust either Fisher estimate?}\label{sec:biasImportance}
We have seen that the combined method is very useful for making inferences on estimated parameter errors, but it is only accurate when the additive bias term on the compressed Fisher estimate is small. How can we estimate if we are in that regime and what qualifies as small? 

Firstly this can be done by applying the standard methods used to assess convergence of the compressed Fisher information. For example varying the number of simulations used and studying the resulting trends. We have seen the shape of these curves, e.g. \cref{fig:constraint_3param_poisson}, depends on the regime. When the biases are small the parameter constraints estimated by the compressed method should change slowly, and decrease, not increase, as more simulations are included in the analysis.

A second test is to use the following approximation for a perturbed matrix inverse
\begin{align} \label{eq:estBiasSize}
{F^\mathrm{comp.}}^{-1}_{ij} &= (\hat{F}^\mathrm{comp.}-\delta F )^{-1}_{ij}  \nonumber \\ 
&\approx {\left.\hat{F}^\mathrm{comp.}\right.}^{-1}_{ij} +{\left.\hat{F}^\mathrm{comp.}\right.}^{-1}_{ik}\delta F_{km} {\left.\hat{F}^\mathrm{comp.}\right.}^{-1}_{mj},
\end{align}
where in the first equality we have rewritten the true compressed Fisher as the estimated Fisher minus the bias terms. Thus we can estimate terms from \cref{eq:expectationFisherCompressed} and use them to compute the second term in \cref{eq:estBiasSize}. If this term is as large as the first term then we are likely in a regime where the bias terms are dominant. Note this test can equally be applied to test whether the standard Fisher estimate is also unbiased. This test is more useful than equivalent tests on the sizes of the biases to the Fisher Information as biases that are small compared to the Fisher information, as in \cref{fig:fisher_3param_poisson}, can still be important for the parameter constraints if there are strong parameter degeneracies.
\section{A test case application}\label{sec:test_case}
As a final, and more complex case, we consider an application to a problem in cosmology. This problem and the details of the application are discussed in our companion papers \citet{Coulton_2022b,Jung_2023a} and \citet{Jung_2023b} and we refer the reader to \citet{Coulton_2022a} and \citet{Coulton_2022b} for more details.

In \citet{Coulton_2022b}, the authors consider how much of the information contained in the distribution of halos, a highly simplified proxy for galaxies in our Universe, can be accessed by measurements of the variance and skewness at different scales. This analysis consists of a Fisher forecast for $8$ parameters -- three characterizing the primary features of interest, called \emph{local}, \emph{equilateral} and \emph{orthogonal} that encode three interesting features potentially present in the very early universe, and four parameters that parameterize the model of our Universe ($h$, $n_s$, $\Omega_m$ and $\sigma_8$) and a nuisance parameter $M_\mathrm{min}$.

The observables are assumed to be well approximated by a Gaussian distribution, which is justifiable by the Central Limit Theorem \citep{Scoccimarro_2000}. Analytically computing the variance and skewness of these cosmological observables is highly challenging due to the non-linear nature of the governing equations. Instead these statistics are typically simulated with expensive simulations, which in \citet{Coulton_2022b} cost approximately $400$ cpu-hours per simulation. The authors generated a large suite of simulations containing 15,000 simulations to compute the covariance matrix and 1000 simulations, per parameter, to compute the derivatives (500 perturbed above and 500 perturbed below the fiducial value to compute a first order central difference).

\begin{figure*}
\centering
 \includegraphics[width=1.\textwidth]{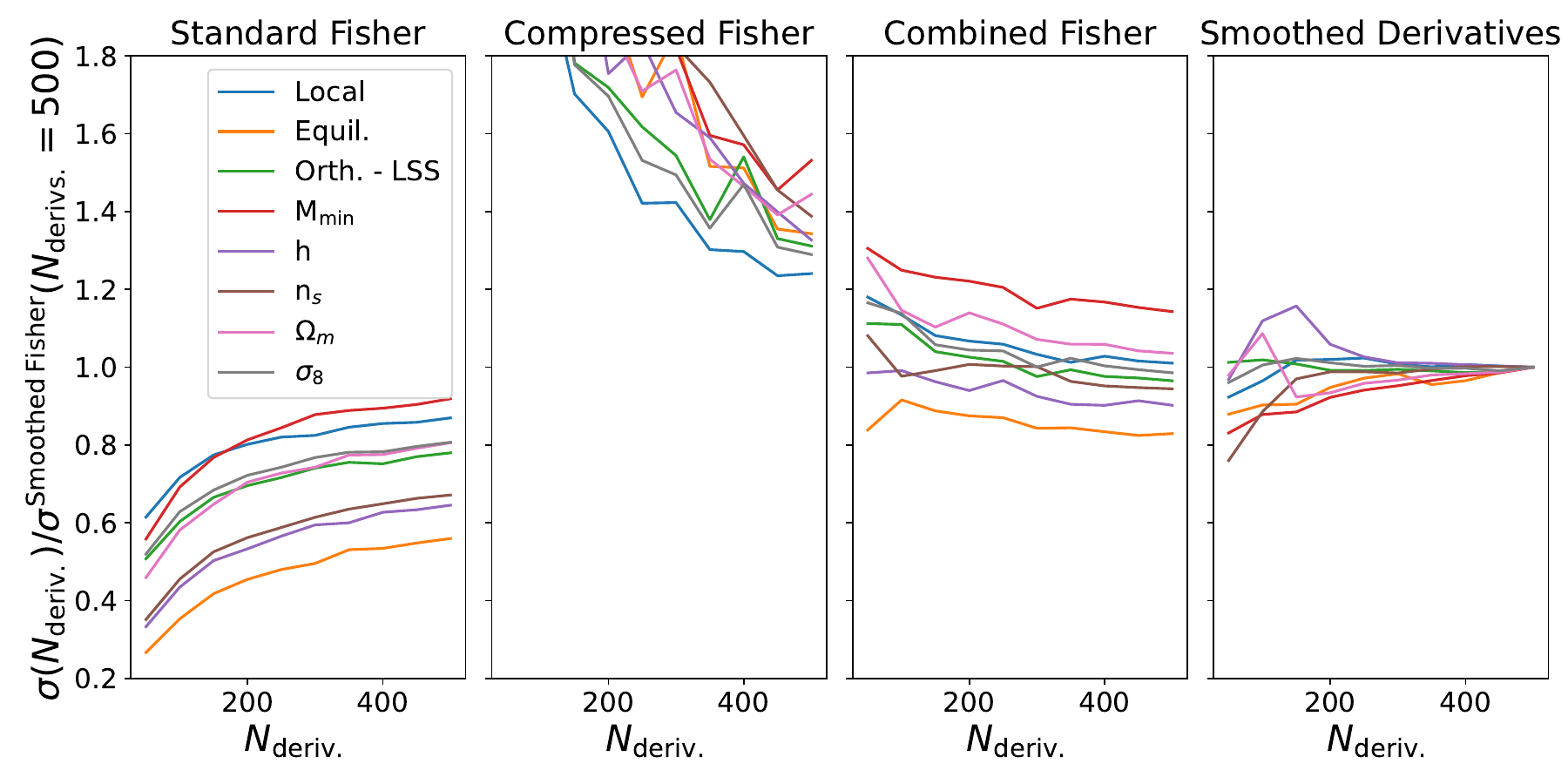}
 \caption{A `real world' test case of our method: a Fisher forecast of what could be learnt from a potential future galaxy survey about three interesting models of the primordial universe, labelled \emph{local}, \emph{equilateral} and \emph{orthogonal}. The analysis jointly accounts for a set of parameters that characterize the key physical processes in our mock observations. We compare four different Fisher forecast methods -- the standard, compressed and combined Fisher forecasts and a Fisher forecast where we have implemented a boutique method to remove noise from the derivatives by carefully smoothing them. \label{fig:testCase}} 
\end{figure*}
The result of the standard Fisher forecast convergence test is show in \cref{fig:testCase}; the forecast errors continue to rapidly increase as the number of simulations are increased implying the results are not converged. Given the high computational cost of each simulation, running a sufficient number is computationally prohibitive. However, the complete data vector of variances and skewness at many physical scales is very large ($\sim 2000$ elements) and thus the compression technique discussed here is potentially very powerful.

\cref{fig:testCase} shows the result of applying the compressed Fisher forecast method. As expected as the number of simulations included in the compressed forecast the constraints decrease -- as the optimality of the compression increases. Remarkably,  the forecast constraints from the combined estimator appear to have converged above $\sim 200$ simulations. We can also use the methods discussed in \cref{sec:biasImportance} to provide confidence that the biases are subdominant, finding that the size of the estimated error ( \cref{eq:estBiasSize}) is subdominant for the compressed estimator, implying that the additive noise bias can be ignored.

To help validate the results \citet{Coulton_2022b} consider an additional method to cross check the results. Given that the biased constraints arise from noise in the derivatives, an alternative method to mitigate the bias would be to fit a smooth function to the derivatives. Generally this is challenging -- if one a priori knew the functional form of the derivatives they would not need to be estimated with simulations! In \citet{Coulton_2022b} the authors use a `blind' method to smooth the noise, whereby the authors fit a Gaussian process to normalized versions of the derivatives \citep[see e.g.,][for a review of Gaussian processes]{RasmussenW06}. The priors chosen for the Gaussian process impose a smoothness on the derivatives and allow the noise, which is highly non-smooth, to be suppressed. Hereafter we refer to this method as `smoothed derivatives'. Fitting functions to smooth the derivatives can be a very successful method to mitigate the noise, but it imposes assumptions on the structure of the derivatives and requires careful and time intensive modeling. By contrast, the method we present in the present paper does not make such assumptions and is therefore generally applicable.

The results from the smoothed case are also shown in \cref{fig:testCase}. The smoothed derivatives show convergence, in this case the smoothing processes is recomputed for each subset of the total derivative data set, and thus we can expect reliable forecasts. We observe a hierarchy of constraints: the forecast errors from the compressed method are larger than the smoothed case, which in turn are larger than the standard case. This ordering is exactly as expected: the standard case is unconverged and so biased to be too small, the compressed case is suboptimal and so biased to be too large. The tailored, smoothed derivatives lie between the two and show generally good agreement with the combined method. Note that even in cases where derivative smoothing is applicable, combining that approach with our convergence-accelerated estimators should still further improve performance. 

\section{Conclusions}\label{sec:conclusions}
In this paper we reviewed the standard practice for performing Fisher forecasts with simulated components. We highlighted how these forecasts can be overly optimistic due to an additive noise bias from Monte-Carlo noise in the simulated derivatives. We then presented two alternative estimators: first the compressed Fisher estimator that can, under certain conditions, provide a conservative estimate of the parameter constraints, giving errors that are biased high due to information lost in the compression. We then showed that the degree to which the standard estimator is biased low is the same as the amount the compressed estimator is biased high. Motivated by this we proposed the second estimator, the combined estimator, that can be used to obtain unbiased estimates of the Fisher information. We provide a public code \textsc{Compressed Fisher }\footnote{\url{https://github.com/wcoulton/CompressedFisher}} that implements these methods.

The key condition for the compressed and combined estimators to be applicable is that an additive noise bias term should be subdominant. Whilst this bias term has the same form as the bias term present for the standard estimator, it is generally smaller by the ratio of number of parameters to the size of the data vector. This ratio can be very significant and enables the compressed and combined estimators to aid simulation-based Fisher forecasts. In \cref{sec:biasImportance} we discuss two tools to assess, for a given use case, if this bias term is subdominant.

Fisher forecasts are a powerful tool used across the physical sciences. With the ever increasing complexity of experiments, it will be increasingly common that the Fisher Information will not be analytically calculable. In those cases, estimating the Fisher Information with simulations will be necessary. The simple tools presented in this paper offer a path to achieving robust simulation-based Fisher forecasts. First these estimators provide a simple test of the standard Fisher forecasts -- if the standard approach is converged we expect the standard and combined estimators to agree. In the case where the standard estimator is unconverged, the combined estimator can accelerate convergence and provide accurate estimates of the Fisher information. This removes the need to run more simulations, often providing effectively unbiased estimates with orders of magnitude fewer simulations, and so dramatically reducing the computational cost. With the rise of automatic differentiation \citep[see e.g.,][]{Neidinger_2010,Baydin_2018}, finite difference derivatives may soon be replaced in many applications. However, automatic differentiation is not without its potential problems  \citep{2023arXiv230503863J}. More generally, the stochasticity of many processes is often not avoided by using such derivatives, and ensemble averages of simulations are still required. Therefore Fisher forecasts with automatic derivatives will suffer the same biases discussed here and can similarly benefit from our new estimators. Likewise, this method can also be combined with other acceleration schemes, such as that discussed in \citet{Chartier_2022}.

We focused our discussion on Fisher forecasts as this is the most likely use case. However, the Fisher information is ubiquitously used across statistics from forming likelihood approximations and confidence intervals to model selections to Bayesian priors \citep[e.g.][]{Rissanen1996FisherIA,Jeffreys_1939,Ly_2017}. Our method can likely also be applied to such applications, if the Fisher information is estimated from simulations. Another interesting application could be in machine learning methods that utilize numerical Fisher estimates, such as \citet{Charnock_2018}.

\section{Data Availability}
A python package implementing these methods is available at https://github.com/wcoulton/CompressedFisher and it includes code to reproduce the toy models discussed here.
\section{Funding}
The Flatiron Institute is supported by the Simons Foundation. 
\bibliographystyle{chicago}
\bibliography{comp_bib.bib}

\end{document}